\documentclass[notitlepage,showkeys,showpacs]{revtex4-1}

\newcommand{\ud}{\mathrm{d}}

\newcommand{\sign}{\mathrm{sign}}
\usepackage{graphicx}
\usepackage{amsmath}
\usepackage{amssymb}
\usepackage{amsfonts}
\usepackage{dsfont}
\usepackage{url}

\begin{document}
\title{Option Pricing Beyond Black-Scholes Based on Double-Fractional Diffusion} 
\author{H. Kleinert}
\email{h.k@fu-berlin.de}
\affiliation{Institute for Theoretical Physics, Freie Universit\"{a}t Berlin, Arnimallee 14, 14195, Berlin, Germany}
\affiliation{ICRANeT Piazzale della Repubblica, 10 -65122, Pescara, Italy}
\author{J. Korbel}
\email{korbeja2@fjfi.cvut.cz}
\affiliation{Faculty of Nuclear Sciences and Physical Engineering, Czech Technical University in Prague, B\v{r}ehov\'{a} 7, 11519, Prague, Czech Republic}
\affiliation{Max-Planck-Institute for the History of Science, Boltzmannstr. 22, 14195 Berlin, Germany}
\date{\today}
\begin{abstract}

We show how the prices of options can be determined with the help of double-fractional differential equation in such a way that their inclusion in a portfolio of stocks provides a more reliable hedge against dramatic price drops than the use of options whose prices were fixed by the Black-Scholes formula.
\end{abstract}
\maketitle 

\section{Introduction}
In 1995, the Nobel prize in Economics was awarded to Merton and Scholes for the so-called Black-Scholes (BS) model
first published in 1973~\cite{Black}. It was hailed as a
milestone in derivative trading, and led to the development of elaborate hedging strategies which promise the safe growth of appropriately composed portfolios of financial assets.
Unfortunately, however, the usefulness of the formula was based on a simplifying assumption
that fluctuations of assets follow Gaussian distributions. This made it possible to set up a mixture of
assets and options that have a chance of growing as a safe investment.
In fact, this formula was initially quite successful to a number of professional speculators. Later, however, it turned out that frivolous trading based only on Black-Scholes model and done without a deeper knowledge of market dynamics can lead to dramatic losses and to the formation of speculative bubbles~\cite{Acharya,Merrouche,Colander}. The reason for this is that rare events such as drastic falls in the stock markets
are much more frequent than would be expected from Gaussian distributions. From time to time catastrophic outlier events
which should not have happened in hundreds of years can occur. Such events have been named "black-swan" events, after a popular book by N.~N.~Taleb~\cite{Taleb}.
The existence of such events is a severe obstacle to all hedging attempts. Normally, one considers price changes as the result of
random steps of a given finite size and demonstrates that these build up to  Gaussian random walks.
This is a consequence of the central limit theorem (CLT).
But in general, some of the steps can be extremely large, and the combined random walks
are of the so-called L\'{e}vy type. These possess power-like tails
which may not even have finite variance. They are encountered in many rare events in nature,
such as earthquakes, monster waves in the ocean, and giant drops in financial markets.
Apparently, we need new option price formulas that incorporate the possibility of encountering large drops in prices,
and are able to compensate for them by a corresponding rise in the price of the derivative.


Many sophisticated models that go beyond Black-Scholes have been introduced in recent decades: among them e.~g.~models based on L\'{e}vy distributions~\cite{Carr},  truncated L\'{e}vy distributions~\cite{KleinertOP}, Multifractal volatility~\cite{Calvet}, jump processes~\cite{Tankov}, and many other approaches. We would like to focus on models that are based on so-called stable distributions. These have found applications in many scientific fields such as multifractal thermodynamics~\cite{Jizba}, quantum field theory~\cite{Swan}, evolutionary systems~\cite{Lee}, complex dynamical systems~\cite{Turcotte}, etc. These models usually exhibit self-similarity and power-law behavior in some particular time interval. We also focus on temporal scaling and show that models based on double-fractional diffusion (i.e., self-similar scaling in both spatial and temporal variables) can successfully simulate situations, when instant fluctuations cause high short-term volatility. In this case we can redistribute the risk for short and long term options  options and the volatility of the model can remain unchanged. This approach has considerable possibilities for usefulness in further applications in more complex option pricing scenarios. By fitting the data of S\&P 500 options, we show that for most of the time the optimal value of temporal-scaling is very close to classical diffusion based on L\'{e}vy flight, but for some particular days, e.g., after sudden drops, the risk redistribution can be accounted for by time-fractional diffusion process.

The paper is divided as follows: in Section~\ref{sec: stable} stable distributions are briefly introduced, in Section~\ref{sec: fractional} we discuss various definitions of fractional derivatives and their relation to stable distributions of the L\'{e}vy type. Section~\ref{sec: option} revises previous models of log-L\'{e}vy option pricing. In Section~\ref{sec: dfdiff} we introduce double-fractional diffusion equations, and discuss their properties. We also show different ways of representing the solutions. Section \ref{sec: dfop} is dedicated to the application of double-fractional diffusion to option pricing and the last section is devoted to conclusions and perspectives.

\section{Stable distributions}\label{sec: stable}
Stable distributions (also called L\'{e}vy distributions) constitute an important class of probability distributions. They are form-invariant under convolution, implying that the sum of two random variables governed by L\'{e}vy distributions follow such a distribution themselves. Gnedenko and Kolmogorov \cite{Gnedenko} showed that such distributions are limiting distributions of infinite sums of independent random variables with arbitrary distribution, which is the content of generalized central limit theorem. Moreover, it is possible to express their probability density function  by Fourier integrals. In probability theory, the logarithms of the Fourier transforms are referred to characteristic functions. In theoretical physics, these  are the well-known Hamiltonian operators $H(p)$. The \emph{stable} Hamiltonian operators can be expressed as a four-parameter set  $H_{\alpha,\beta;\bar{x},\bar{\sigma}}(p)$ of the following form
\begin{equation}
\label{eq: stablecrit}
H_{\alpha,\beta;\bar{x},\bar{\sigma}}(p) \equiv \ln \langle e^{ipx} \rangle = i \bar{x} p - \bar{\sigma}^\alpha |p|^\alpha \left(1 - i\beta \sign(p) \omega (p,\alpha) \right)\, ,
\end{equation}
where
\begin{equation*}
\omega(p,\alpha) =  \left\{ \begin{array}{l}\tan(\pi \alpha/2) \quad \mathrm{if}  \quad \alpha \neq 1,
\\ \frac{2}{\pi} \ln |p| \quad \mathrm{if} \quad \alpha = 1. \end{array} \right.
\end{equation*}
Each stable hamiltonian corresponds to the stable distribution, which is denoted as $L_{\alpha,\beta;\bar{x},\bar{\sigma}}(x)$. The parameters acquire following values:  $\alpha \in (0,2]$, $\beta \in [-1,1]$, $\bar{\sigma} \geq 0$, and $\bar{x} \in \mathds{R}$. Unless specified differently, we consider the standard case, i.e., $\bar{x}=0$ and $\bar{\sigma} = 1$, and denote the stable Hamiltonian as $H_{\alpha,\beta}(p)$.

For $\alpha = 2$, regardless of $\beta$, the distribution $L_{2,\beta}(x)$ is simply the Gaussian distribution. The parameter $\bar{x}$ has an interpretation of a location parameter, and for $\alpha > 1$ is equal to the expectation value $\langle x \rangle$. Parameter $\bar{\sigma}$ is a scaling parameter, for $\alpha=2$ equal to the standard deviation. Parameter $\beta$ is an \emph{asymmetry parameter} and the parameter $\alpha$ is called  \emph{stability parameter} and determines the overall behavior of the probability distribution function. For $\alpha < 2$ the L\'{e}vy distribution $L_{\alpha,\beta}(x)$ exhibits power-like L\'{e}vy tails that behave like $x^{-(\alpha+1)}$ for positive $x$ (see Ref.~\cite{Kleinert}), so
\begin{equation}
L_{\alpha,\beta}(x) \sim \alpha  C_\alpha (1+\beta) x^{-(\alpha+1)} \quad \mathrm{for} \ x \rightarrow +\infty
\end{equation}
with a similar behavior for negative $x \rightarrow -\infty$. The only exception is the case of  $\beta = -1$, where for $\alpha > 1$ exhibits the distribution the following behavior
\begin{equation}
L_{\alpha,-1}(x)   \sim \frac{1}{2(\alpha-1)} \left(\frac{x}{\alpha c_\alpha}\right)^{\frac{\alpha}{2(\alpha-1)}-1} \exp\left[ -(\alpha-1)\left(\frac{x}{\alpha c_\alpha}\right)^{-\frac{\alpha}{(\alpha-1)}} \right]\, \quad \mathrm{for} \ x \rightarrow +\infty
\end{equation}
with some constant $c_\alpha$. The proof can be found in \cite{Zolotarev}. A similar exponential decay holds for the negative tail and $\beta = 1$. For $\alpha \leq 1$, the support  is bounded to the interval $[\bar{x}, \infty)$ for $\beta = 1$, resp. $(-\infty,\bar{x}]$ for $\beta = -1$. For $\alpha < 2$, the distribution has infinite moments $\langle|x|^l\rangle$ for $l \geq \alpha$.

The two-sided Laplace transform of $L_{\alpha,\beta}$ does not exist, unless $\beta=1$, so for $\Re(\lambda) > 0$ (see Ref. \cite{Zolotarev}) the logarithm of Laplace image is equal to
\begin{equation}
\label{eq: laplacelevy}
\ln\langle e^{-\lambda x}\rangle = -\lambda \bar{x} - \lambda^\alpha \bar{\sigma}^\alpha \sec \frac{\pi \alpha}{2}.
\end{equation}
From the symmetry argument we can deduce that the expectation value $\langle e^{\tau x} \rangle$ for $\tau \geq 0$ exists only when $\beta = -1$.
Sometimes, it is advantageous to use an alternative representation of the stable Hamiltonian, which is
\begin{equation}\label{eq: stablediamon}
\mathcal{H}_{\alpha,\theta;\bar{x},c}(p) \equiv \ln \langle e^{ipx} \rangle =  i \bar{x} p - c | p|^\alpha e^{i \, \sign(p) \, \theta \, \frac{\pi}{2}}\, ,
\end{equation}
where $c$ and $\theta$ are uniquely determined by the parameters $\alpha$, $\beta$ and $\sigma$~\cite{Sato}. Accessible values of the parameter $\theta$ satisfy condition  $|\theta| \leq \min \{\alpha,2-\alpha\}$. The corresponding area in $(\alpha,\theta)$-plane is called Feller-Takayasu diamond, and for certain values we obtain extremely asymmetric stable distributions (e.g., for $\alpha >1$, the value $\theta = -\alpha+2$ corresponds to the case $\beta = 1$ and $\theta = \alpha-2$ corresponds to the case $\beta=-1$). Further properties of stable distributions, their representations and relations between them can be found in Refs.~\cite{Kleinert,Samoradnitsky, Sato}.


\section{Fractional calculus}\label{sec: fractional}
Fractional calculus generalizes the classical integral and differential calculus to non-integer integration and differentiation. The motivation was originally to find such an operator, for which the second power of the operator would be equal to the first derivative. Subsequently, the goal was to generalize the integral and differential calculus for all real values. This section presents a basic overview of some possible definitions.

Let us begin with the fractional integration, which generalizes the well-known Cauchy formula
\begin{eqnarray*}
\int_{x_0}^x \int_{x_0}^{x_1}\dots \int_{x_0}^{x_{n-1}} f(x_n) \ud x_n \dots \ud x_1  = \frac{1}{(n-1)!}\int_{x_0}^x (x-y)^{n-1} f(y) \ud y \,
\end{eqnarray*}
in a natural way from integer values $n$ to non-integer values $\nu$ by defining
\begin{equation}{}_{x_0}\mathcal{I}^{\nu}_{x} f(x) := \frac{1}{\Gamma(\nu)}\int_{x_0}^x (x-y)^{\nu-1} f(y) \ud y.\end{equation}

The fractional integral is a linear operator that obeys the semigroup property
\begin{equation}{}_{x_0}\mathcal{I}^{\nu_1}_{x} \circ {}_{x_0}\mathcal{I}^{\nu_2}_{x} = {}_{x_0}\mathcal{I}^{\nu_1+\nu_2}_{x}.\label{eq: semigroup}\end{equation}
It is easy to show that $\frac{\ud}{\ud x} \left( {}_{x_0}\mathcal{I}^{\nu+1}_x \right) = {}_{x_0}\mathcal{I}^{\nu}_x$, which is the baseline for the definition of fractional derivative.
\subsection{Riemann-Liouville derivative}
The relation between fractional integrals and derivatives suggests the introduction the Riemann-Liouville (RL) fractional derivative
\begin{equation}\label{eq: rlder}
 {}_{x_0}\mathcal{D}_{x}^\nu f(x) := \frac{\ud ^{\lceil \nu \rceil }} {\ \ud ^{\lceil \nu \rceil} x}  \left( {}_{x_0}\mathcal{I}_{x}^{ \lceil \nu \rceil - \nu}[f]\right)(x)\, ,\end{equation}
 where $\lceil \nu \rceil$ denotes the associated ceiling function, i.e., the lowest integer exceeding $\nu$. As with ordinary derivatives, these derivatives are linear operators, but contrary to the former, they do not
satisfy a composition law like~\eqref{eq: semigroup} , i.e.,
\begin{equation}
{}_{x_0}\mathcal{D}_{x}^{\nu_1}\circ {}_{x_0}\mathcal{D}_{x}^{\nu_2} \neq {}_{x_0}\mathcal{D}_{x}^{\nu_2} \circ {}_{x_0}\mathcal{D}_{x}^{\nu_1}.
\end{equation}
In fact, they share only a few of the properties of ordinary derivative operators. On the other hand, some particular choices of derivatives, given by special values of $x_0$, recover some of the typical properties of ordinary derivatives. One example of such a fractional derivative can be inferred from the derivative of integer power functions $x^n$. For arbitrary integers $m,n$ is the $m$-th derivative of $x^n$ equal to
 \begin{equation}\label{eq: mon}
 \frac{\ud^m}{\ud x^m} x^n = \frac{n!}{(n-m)!}x^{n-m}\, .
 \end{equation}
By performing a fractional integration, it is easy to show that the fractional derivative of a monomial has the same form as in Eq.~\eqref{eq: mon}, when $x_0 = 0$. This operator will be denoted  $\mathcal{D}_{x}^\nu \equiv {}_{0}\mathcal{D}_{x}^\nu$. Moreover, this property is valid for arbitrary powers and arbitrary derivatives. As an example, we obtain for $n=0$ that
\begin{equation}\label{eq: der}
\mathcal{D}_{x}^\nu 1 = \frac{x^{-\nu}}{\Gamma(1-\nu)}\, .
\end{equation}
Expression~\eqref{eq: der} is equal to zero only for $\nu \in \mathds{N}$, because of poles in the Gamma function.

Sometimes, it is advantageous to discuss the formalism at the level of the Laplace images. If we perform the Laplace transform of Def.~\eqref{eq: rlder}, we get
\begin{equation}
\mathcal{L}\left[\mathcal{D}_{x}^\nu f(x);s\right] \equiv \widehat{[\mathcal{D}_{x}^\nu f]}(s) = s^\alpha F(s) - \sum_{k=0}^{\lfloor \nu \rfloor} s^k \left[\mathcal{D}_x^{\nu-k-1} f(x) \right]_{x=0}\, ,
\end{equation}
where $\lfloor x \rfloor$ represents floor function, i.e., the largest integer not exceeding $x$. More details can be found in Ref.~\cite{Podlubny}.
\subsection{Caputo fractional derivative}
One has to notice that the Riemann-Liouville definition of fractional derivatives has some objectionable properties, among them non-zero derivative of constant function and unnatural initial conditions. The latter issue for the most part limits the application of the above derivative in physical and other real problems, because the fractional initial conditions have no physical meaning. These reasons compel us to introduce a different kind of fractional derivative, namely the Caputo (C) fractional derivative, which patches up some of the unwanted properties of RL derivatives. It is defined as follows
\begin{equation}\label{eq: caputoder}
{}_{x_0}{}^\ast\mathcal{D}_{x}^\nu f(x) := \frac{1}{\Gamma(\lceil \nu \rceil - \nu)} \int_{x_0}^x \frac{f^{\lceil \nu \rceil}(y)}{(x-y)^{\nu+1-\lceil \nu \rceil}} \ud y\, .
\end{equation}
The Caputo derivative is more restrictive regarding its domain, because the function $f$ must have at least $\lceil \nu \rceil$ derivatives. On the other hand, it recovers the desired properties, because ${}^\ast\mathcal{D}_{x}^\nu 1 = 0$, and the Laplace transform is equal to
\begin{equation}
\mathcal{L}\left[{}^\ast\mathcal{D}_{x}^\nu f(x);s\right] \equiv \widehat{[{}^\ast\mathcal{D}_{x}^\nu f]}(s) = s^\nu F(s) - \sum_{k=0}^{\lfloor \nu \rfloor} s^{\nu-k-1} f^{(k)}(0)\, .
\end{equation}
 Now, natural initial conditions are recovered. In case of the Caputo derivative, we can solve the fractional differential equation
\begin{equation}\label{eq: caputoeigen}
{}^\ast\mathcal{D}_{x}^\nu f(x) = \lambda f(x)\,
\end{equation}
by introduction of the Mittag-Leffler function
\begin{equation}
E_{\nu,\zeta}(x) = \sum_{n=0}^\infty \frac{x^n}{\Gamma(\nu n +\zeta)}\, .
\end{equation}
It is easy to see that the function $f(x) = E_{\nu,1}(\lambda x^\nu)$ solves Eq. \eqref{eq: caputoeigen}.
Eventually, Riemann-Liouville and Caputo derivatives can be connected via the relation (to be found e.g. in Ref.~\cite{Abdeljawad})
\begin{equation}
{}^\ast\mathcal{D}_{x}^\nu f(x) = \mathcal{D}_{x}^\nu f(x) - \sum_{k=0}^{\lfloor \nu \rfloor} \frac{x^k}{k!}f^{(k)}(0)\, .
\end{equation}
\subsection{Riesz-Feller fractional derivative}
Equation \eqref{eq: caputoeigen} is a generalization of another well-known property of derivatives, which holds for exponential functions
\begin{equation}\label{eq: rieszeigen}
\frac{\ud^m}{\ud x^m} \exp(\lambda x) = \lambda^m \exp(\lambda x)\, .
\end{equation}
In the case of the Caputo derivative, this formula is generalized in a way that the exponential function is replaced by the Mittag-Leffler function. On the other hand, this property can be preserved, when we send the lower bound to minus infinity; the operator
\begin{equation}\label{eq: rieszder}
D_{x}^\nu f(x):= \lim\limits_{x_0 \rightarrow -\infty}{}_{x_0}\mathcal{D}_{x}^\nu f(x)
\end{equation}
obeys Eq.~\eqref{eq: rieszeigen} even for non-natural derivatives.  Such an operator has to be defined on a different space, because the fractional operator of this type is convergent for functions that decay faster than $|x|^{-\nu}$ for $x \rightarrow -\infty$.  This operator is called the Riesz-Feller derivative and is denoted by $D_x^\nu$. Interestingly, we get the same operator, when we use the Caputo derivative approach. In the case of Riesz-Feller derivative it is advantageous to transform into Fourier image because it has the following form
\begin{equation}
\mathcal{F}\left[D^\nu f(x);p \right] \equiv \overline{[D^\nu f]}(p)= \int_\mathds{R} \ud x e^{i p x} \int_{-\infty}^x \ud y (x-y)^{-\nu-1} f(y) = (-ip)^{\alpha} \overline{f}(p).
\end{equation}
The proof can be found also in \cite{Samko}. Actually, the Riesz-Feller derivative operator is a special case of the so-called L\'{e}vy pseudo-differential operators, for which the fractional Laplacian $(-\Delta)^{\nu/2}$ is an important example. This class of pseudo-differential operators is defined via the Fourier transform through a Hamiltonian
\begin{equation}
\overline{\left[{}^\theta D^\nu_x f\right]}(p) = \mathcal{H}_{\nu,\theta}(p) \overline{f}(p)\, ,
\end{equation}
where $\mathcal{H}_{\nu,\theta}(p)$ is the function defined in Eq.~\eqref{eq: stablediamon}. Particularly, for extreme values, i.e., for $|\theta| = \min\{\nu,2-\nu\}$ we obtain operators proportional to Riesz-Feller fractional derivatives. For the choice $\theta=0$, we have the fractional Laplacian. These differential operators are closely related to L\'{e}vy distributions and L\'{e}vy processes, because they are the solutions for the space-fractional diffusion equation
\begin{equation}
\frac{\partial}{\partial t} f(x,t) = {}^\theta D^\nu_x f(x,t)\, .
\end{equation}
\section{Option pricing under Log-L\'{e}vy process}\label{sec: option}
In classic Black-Scholes option pricing theory \cite{Black}, the price of an option is calculated with the assumption of an efficient market. The evolution of an underlying asset is modeled as a log-normal (or geometric Brownian) process
\begin{equation}
\ud S(t) = \mu_S S(t) \ud t + \sigma_S S(t) \ud W(t)\, ,
\end{equation}
where $W(t)$ is a standard Wiener process, $\mu_S$ is the drift, and $\sigma_S$ is the volatility of the process. In the following, we will be using notation $S_t \equiv S(t)$. The assumption of an efficient market implies that there exists another probability measure $\mathbb{Q}$, equivalent to the real probability measure $\mathbb{P}$, under which the discounted price is a martingale, so $S_t = e^{-\mu_S(T-t)} \langle S_T|\mathcal{F}_t\rangle_\mathbb{Q}$, where $\mathcal{F}_t$ is a filtration of the stochastic process at time $t$, and $T$ is the maturity time. For the exponential processes, the change of measures is given by Esscher's transform \cite{Gerber}, which states that the Radon-Nikodym derivative is
\begin{equation}
\frac{\ud \mathbb{Q}_{\mathcal{F}_t}}{\ud \mathbb{P}_{\mathcal{F}_t}} = \frac{e^{X_t}}{\langle e^{X_t}\rangle_\mathbb{P}} = \exp\left[X_t - \mu_\mathbb{Q} t\right]
\end{equation}
where $\mu_\mathbb{Q} =  \ln \langle \exp(X_1)\rangle$.
Particularly, in case of log-normal distribution with the drift equal to interest rate $r$, the process of price evolution has the following form
\begin{equation}
S_t = S_0 \exp\left[\left(r-\frac{\sigma^2}{2}\right)t  + \sigma W(t)\right].
\end{equation}
The price of the option $H$ can therefore be calculated as \cite{Tankov}
\begin{equation}\label{eq: optionprice}
H(t) = e^{-r(T-t)}\langle H(T)|\mathcal{F}_t\rangle_\mathbb{Q}.
\end{equation}
For the European call option, the pricing formula is, with the notation $[x]^+ = \max\{x,0\}$, equal to
\begin{equation}
C(S_t,t) = e^{-r\tau} \int_\mathds{R} \ud x \left[S_t e^{r \tau  + x}-K\right]^+  \frac{1}{\sqrt{2\pi \tau \sigma^2}} e^{-\frac{\left(x-\frac{\sigma^2}{2}\right)^2}{2\tau\sigma^2}}.
\end{equation}
By direct differentiation it is easy to show that the option this price fulfills the celebrated Black-Scholes equation
\begin{equation}
\frac{\partial C(S,t)}{\partial t} =  r C(S,t) - r S \frac{\partial C(S,t)}{\partial S} - \frac{1}{2} \sigma^2 S^2 \frac{\partial^2 C(S,t)}{\partial S^2}.
\end{equation}
with the boundary condition $C(S_T,T) = \left[S_T-K\right]^+$.

This option pricing model is the one most used worldwide, with many applications, such as  estimation of implied volatility~\cite{Stanley}. On the other hand, a considerable amount of effort has been made during last two decades on the development of more advanced evolution models of the underlying assets. As an example, the fractional Brownian motion~\cite{Mandelbrot} is an elegant model. Furthermore, the complexity of financial markets motivated many authors to introduce even more sophisticated option pricing rules, some examples are provided by Refs.~\cite{Heston, Necula, Gerber}.

We adopt the approach introduced in Ref.~\cite{Carr} and assume that the price evolution is driven by the log-L\'{e}vy model
\begin{equation}
\ud S(t) = r S(t) \ud t + \sigma  S(t) \ud L_{\alpha,\beta}(t).
\end{equation}
Contrary to geometric Brownian process, the prices of log-L\'{e}vy stable process do not possess all moments $\langle S_t^n \rangle$. What is more, it is not possible to use Esscher transform. The only exception is the case, when $\beta=-1$, or similarly $\theta = \alpha-2$, because then the two-sided Laplace transform  exists, which is expressed in~Eq. \eqref{eq: laplacelevy}. As a result of the exponential decay of the positive tail of the PDF, all moments exist and are finite. Such a model can better describe dramatic price drops on the market, which are more frequent than was envisaged by Black-Scholes
theory \cite{Swan}.  The price process becomes the following time dependence
\begin{equation}
S_t = S_0 \exp\left[\left(r + \mu \right)t  + \sigma L_{\alpha,-1}(t)\right]\, ,
\end{equation}
where $\mu = \sigma^\alpha \sec \frac{\pi \alpha}{2}$. The corresponding option price, which is given by \eqref{eq: optionprice}, is equal to
\begin{eqnarray}
C(S_t,t) = e^{-r\tau} \int_\mathds{R} \ud x \left[S_t e^{r \tau  + x}-K\right]^+ \int_\mathds{R} \ud k \frac{e^{-i p x}}{2 \pi} e^{\tau \left[i p \mu + \sigma^\alpha H_{\alpha,-1}(p) \right]}\, ,
\end{eqnarray}
where $\tau = T-t$. By introduction of a new variable $z = \ln S_t$ and change of integration variable to $y = x + r \tau + z$, we obtain
\begin{eqnarray}
C(z,\tau) = e^{-r \tau} \int_\mathds{R} \ud y \left[e^y - K\right]^+ \int_\mathds{R} \frac{\ud p}{2 \pi} e^{i p (z-y)} e^{\tau \left[ ip(r+\mu) - \mu (i p)^\alpha \right]} =
\int_\mathds{R} \ud y \left[e^y - K\right]^+ \tilde{g}_\alpha(z,\tau|y,0)\, ,
\end{eqnarray}
where $\tilde{g}_\alpha(z,\tau|y,0)$ is the Green function (sometimes also called fundamental solution). By further transformations $\xi = z - y + \tau(r + \mu)$ and $g_\alpha(\xi,\tau) = e^{r \tau} \tilde{g}_\alpha(z,\tau|y,0)$ we obtain the equation for $g_\alpha$ in the form
\begin{equation}
\frac{\partial g_\alpha(\xi,\tau)}{\partial \tau} = -\mu \left[{}^{\alpha-2}D^\alpha_\xi g_\alpha\right](\xi,\tau)\, ,
 \end{equation}
together with the initial condition $g_\alpha(\xi,0) = \delta(\xi)$. This equation is a fractional Black-Scholes equation for log-prices, and for $\alpha=2$, we recover the classical diffusion equation. In the Fourier image, the equation has the form of a fractional-diffusion equation
\begin{equation}
\frac{\partial \bar{g}_\alpha(p,\tau)}{\partial \tau} =  H_{\alpha,-1}(p) \bar{g}_\alpha(p,\tau).
\end{equation}
\section{Double-fractional diffusion}\label{sec: dfdiff}
In some recent works, other models that involve fractional time derivatives have been studied~\cite{Jumarie,Song,Wyss}. It has been discovered that complex time scaling introduces more general classes of solutions that exhibit interesting phenomena and enables one to price options more realistically. The original motivation was provided by fractional Brownian motion, originally introduced in~\cite{Mandelbrot}, and later applied to option pricing in Refs.~\cite{Hu,Necula}. The question at stake is which particular fractional derivative is the best when generalizing the Black-Scholes equation, driven by a Green function $g(\xi,\tau)$, obtained as a solution of a \emph{double-fractional diffusion equation}
\begin{equation}
\left({}^K\!\partial^\gamma_\tau + \mu  [{}^{\alpha-2}D^\alpha_\xi] \right) g(\xi,\tau) = 0\, ,
\end{equation}

where we have considered two type of the temporal derivatives denoted by the parameter $K$, namely ${}^C\!\partial^\gamma = {}^\ast\mathcal{D}^\gamma$ (Caputo derivative), or ${}^{RF}\!\partial^\gamma = D^\gamma$ (Riesz-Feller derivative). The parameter $\alpha$ is the degree of the spatial derivative and corresponds to the stability parameter. The parameter $\gamma$ is the degree of the temporal fractional derivative (corresponding to parameter $\nu$ in definitions of fractional derivatives). It is called the \emph{diffusion speed parameter}, because, as discussed below, it influences the speed and type of diffusion behavior. In the following sections, we compare two classes of double-fractional diffusion equations, particularly equations with Riesz-Feller time derivatives, resp. Caputo time derivatives. Both of these equations are examples of a wide class of two-variable pseudo-differential equations, which are usually represented via the Laplace-Fourier (LF) image (which means the Fourier image in spatial variable and the Laplace image in the temporal variable) as

\begin{equation}
\label{eq: doublefract}
a(s) \hat{\bar{g}}(p,s) - a_0(s) \bar{g}_0(p) \ = \ b(p) \hat{\bar{g}}(p,s)\, \,
\end{equation}

In our case, i.e. when Eq.~\eqref{eq: doublefract} represents a LF image of the double-fractional diffusion equation, then \mbox{$b_\alpha(p) = H_{\alpha,-1}(p)$}, $a_\gamma(s)= s^\gamma$ and $a_0(s)$ is determined by the type of derivative, for the Caputo derivative $a_0^C(s) = s^{\gamma-1}$, and for Riesz-Feller derivative is $a_0^{RF}(s) = 1$. We shall note that the equation requires one initial condition $\bar{g}_0(p)$, which is the Fourier transform of $g_0(\xi) \equiv g(\xi, \tau=0)$. In the following, we consider that $g_0(\xi) = \delta(\xi)$, which leads to $\bar{g}_0(p) \equiv 1$. This is valid for the case, when $\gamma < 1$ (sometimes called \emph{slow diffusion}). In the following section we focus on that case, because there exists a representation which results in a nice interpretation of double fractional diffusion. On the other hand, when we consider $ 1< \gamma \leq 2$ (\emph{fast diffusion}), we can also obtain the same form of equation as in Eq.~\eqref{eq: doublefract}, we have only to add the second initial condition of the particular form, namely

\begin{equation}
\frac{\partial g}{\partial \tau}(\xi, \tau = 0) \equiv 0\, .
\end{equation}

The question at stake is, whether the solutions of Eq.~\eqref{eq: doublefract} can be interpreted as probability distributions, i.e. whether they are positive. In Ref.~\cite{Mainardi} it is shown that for $\gamma <1$ are solutions positive for all $\alpha \in (0,2]$, while in case $\gamma > 1$, the parameters have to obey the condition $0 < \gamma < \alpha \leq 2$.

\subsection{Composition rule for $\gamma<1$ and smearing kernel}
In the case when $\gamma <1$, we can derive a special form of the Green function, which gives a nice interpretation of the double fractional Green function as a composition of space-fractional Green functions weighted by smearing kernel equal to time-fractional Green function. We begin with Eq.~\eqref{eq: doublefract}. The solution to this equation can, in Fourier-Laplace image, be expressed as

\begin{equation}
\hat{\bar{g}}(p,s) = \frac{a_0(s) \bar{g}_0(p)}{a(s)-b(p)}.
\end{equation}

Under the assumption that $\Re(a(s)+b(p))>0$, we can apply Schwinger's formula and obtain
\begin{equation}
\hat{\bar{g}}(p,s) = \int_0^\infty \ud l \, a_0(s) e^{-l a_\gamma(s)} \bar{g}_0(p) e^{l b_\alpha(p)} = \int_0^\infty \ud l \, \hat{g}_S(s,l) \bar{g}_0(l,p)\,.
\end{equation}
Therefore, the original $(\xi,\tau)$-dependence is given by
\begin{equation}
g(\xi,\tau) = \int_0^\infty \ud l\, g_K(\tau,l) g_0(l,\xi)\, .
\end{equation}
Because $\int_\mathds{R} \, g(\xi,t) \ud \xi = 1$, we obtain that $\int_0^\infty g_K(\tau,l) \ud l = 1$. One possible interpretation of the variable $l$ is that it represents the generic time-parameter of the system and the function $g_1(\tau,l)$ represents the smearing kernel, so that the resultant solution is obtained by integration over all solutions for different time-parameters with the weight factor given by the smearing kernel.

After plugging in for $b(p)$ and $g_0(p)$, we obtain the solution in the form
\begin{equation}
\hat{\bar{g}}(p,s) = \int_0^\infty \ud l \, a^0(s) e^{-l a(s)}  e^{l \cdot H_{\alpha,-1}(p)}\, = \int_0^\infty \hat{g}_K(s,l) \bar{g}_\alpha(l,p) ,
\end{equation}
so the solution is given by the superposition of space-fractional-diffusion Green functions for different times $l$. The smearing kernel $g_K(s,l)$  obeys the differential equation
\begin{equation}
\frac{\ud}{\ud l} \hat{g}_K(s,l) \ = \ - a(s) \hat{g}_K(s,l)
\end{equation}
with an initial condition $\hat{g}_K(s,0) = a_0(s)$.

\textbf{Riesz-Feller kernel:} when we assume that the time-derivative operator is equal the Riesz fractional derivative $D^\gamma_\tau$ this results in the expected property that for $\tau < 0$ is $g_K(\tau,l) \equiv 0$~\cite{Zatloukal}. We find that the Laplace image is equal to~\cite{Zhang}:
\begin{equation}
\frac{\ud}{\ud l} \hat{g}_K^{RF}(s,l) = -s^\gamma \hat{g}_K^{RF}(s,l)
\end{equation}
with the initial condition $\hat{g}_K^{RF}(s,0) = 1$, leading to the solution $\hat{g}_K^{RF}(s,l) = e^{-l s^\gamma}$, which nothing else than the Laplace transform of the fully asymmetric stable distribution with stability parameter $\gamma$, asymmetry parameter equal to 1 and the support contained in $\mathds{R}_0^+$. The function $g_K^{RF}(t,l)$ is not normalized, so according to~\cite{Zatloukal}, we have
\begin{equation}\label{eq: riesz}
\int_0^\infty \ud l\, g_K^{RF}(\tau,l) = \int_0^\infty \ud l \, \int_\mathds{R} \ud p\, \frac{e^{-i p \tau}}{2\pi} e^{-l \mu(i p)^\gamma}
= \int_\mathds{R} \ud p \frac{e^{-i p \tau}}{2\pi} \frac{1}{\mu (i p)^\gamma} = \frac{\tau^{\gamma-1}}{\mu \Gamma(\gamma)}.
\end{equation}
Thus, we end with
\begin{equation}
g^{RF}(\xi,\tau) = \int_0^\infty \ud l\, \left(\frac{\Gamma(\gamma)}{\tau^{\gamma-1}}\right) \frac{1}{l^{1/\gamma}} L_{\gamma,1}\left(\frac{\tau}{l^{1/\gamma}}\right) g_{\alpha}(\xi,l)
\end{equation}
where $L_{\gamma,1}$ is the $\gamma$-stable L\'{e}vy asymmetric distribution.

\textbf{Caputo kernel:} in case of Caputo fractional derivative ${}^*\mathcal{D}^\gamma_\tau$, the Laplace transform of $\hat{g}^C_K(s,l)$ is equal to $\hat{g}_K(s,l) = s^{\gamma-1} e^{-l s^\gamma}$. According to Ref.~\cite{Gorenflo2}, the inverse Laplace transform is equal to
\begin{equation}
g_K^C(\tau,l) = \frac{1}{\tau^\gamma} M_\gamma\left(\frac{l}{\tau^\gamma}\right)
\end{equation}
where $M_\nu(z)$ is the $M$ function of Wright type, which is defined by the infinite series
\begin{equation}
M_\nu(z) = \sum_{n=0}^\infty \frac{(-z)^n}{n! \Gamma(-\nu n + (1-\nu))}\, .
\end{equation}
Interestingly, the connection of $M$-function to asymmetric L\'{e}vy-distributions is provided by the relation
\begin{equation}
\frac{1}{c^{1/\nu}} L_{\nu,1}\left(\frac{x}{c^{1/\nu}}\right) = \frac{c \nu}{x^{\nu+1}} M_\nu \left( \frac{c}{x^\nu}\right)\, ,
\end{equation}
which is valid for $\nu \in (0,1)$, $c>0$ and $x>0$.
%
%
Hence, we can rewrite the Green function $g^C(\xi,t)$ as
\begin{equation}
g^C(\xi,\tau) = \int_0^\infty \ud l \, \frac{1}{\tau^\gamma} M_\gamma\left(\frac{l}{\tau^\gamma}\right) g_{\alpha}(\xi,l) = \int_0^\infty \ud l \left(\frac{\tau}{l \gamma}\right) \frac{1}{l^{1/\gamma}}L_{\gamma,1}\left(\frac{\tau}{l^{1/\gamma}}\right) g_\alpha(\xi,l)\, .
\end{equation}
We compare the properties of both smearing kernels in Appendix~\ref{sec: appendix}. Now let us turn attention to an alternative representation of the Green function, which is more computationally tractable and includes $\gamma > 1$.

\textbf{Representation of expectations:} The composite representation can be also used in the case of calculating expectation values
\begin{equation}
<F(X)>_{DF} = \int_\mathds{R} F(\xi) g^{DF}(\xi,\tau) \ud \xi\, .
\end{equation}
Because the composition rule can also be formally used  for $\gamma>1$, we can rewrite the previous expression as
\begin{equation}
\int_\mathds{R} F(\xi) g^{DF}(\xi,\tau) \ud \xi\, = \int_\mathds{R} F(\xi) \int_0^\infty \ud l\, g_K(\tau,l) g_0(l,\xi)\, = \int_0^\infty \ud l\, g_K(\tau,l) <F(X)>_{L_{\alpha}(l)}
\end{equation}
where $<F(X)>_{L_{\alpha}(l)}$ is the expectation value under L\'{e}vy-stable process in pseudo-time $l$. This is important in the calculation of Risk-neutral measure by Esscher's transform
\begin{equation}
\mu_{DF} = \ln \langle \exp(X_1) \rangle_{DF} = \ln \int_\mathds{R} \exp(\xi) \int_0^\infty \ud l\, g_K(1,l) g_0(l,\xi) = \int_0^\infty \ud l\, g_K(1,l) \exp\left(-l^\alpha \sec \frac{\alpha \pi}{2}\right)\, .
\end{equation}
\subsection{Mellin-Barnes representation of double-fractional-diffusion Green function}
%
\begin{figure}[t]
\begin{center}
\includegraphics[width=6cm]{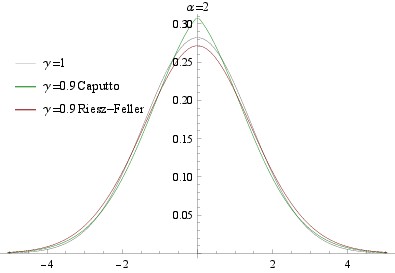} \quad
\includegraphics[width=6cm]{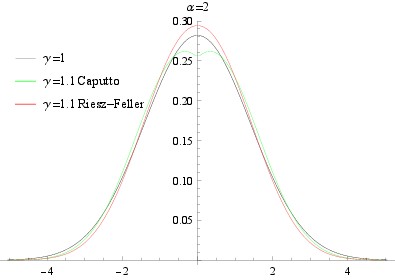}\\
\includegraphics[width=6cm]{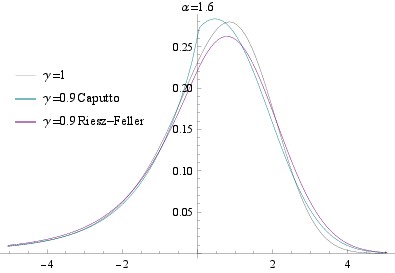} \quad
\includegraphics[width=6cm]{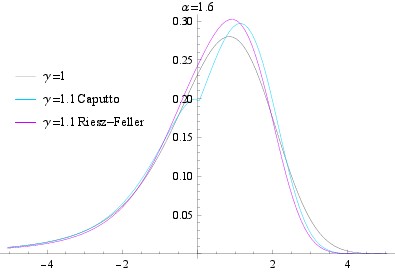}
\end{center}
\caption{Comparison of Green functions for ordinary derivative ($\gamma=1$), Riesz and Caputo derivative for $\gamma=0.9$ (slow diffusion) and $\gamma=1.1$ (fast diffusion) for $\alpha=2$ and $\alpha=1.6$ . The Caputo Green function highlights the peak of the distribution, while Riesz-Feller Green function has slower decay in tails of the distribution. Note that for $\gamma > 1$, the green function exhibits wave-like behavior with two peaks receding in time.}
\label{fig: green}
\end{figure}
It is possible to introduce an alternative, computationally effective representation of double-fractional diffusion equation which is based on the Mellin-Barnes~\cite{Paris} transform. This representation is also valid for the case $\gamma >1$. Let us again begin with equation \eqref{eq: doublefract}. It has the solution equal to

\begin{equation}\label{eq: diffrepr}
\hat{\bar{g}}(p,s) = \frac{s^{\gamma-\kappa}}{s^\gamma - \mathcal{H}_{\alpha,\theta}(p)},
\end{equation}
where $\kappa$ depends on the type of the derivative. For Riesz derivative is $\kappa=\gamma$, for Caputo derivative we have $\kappa=1$. We shall note that because of computational reasons, we assume here only solutions for $\xi>0$. The solution for negative values can be then easily obtained from the relation $g_{\alpha,\theta,\gamma,\kappa}(\xi) = g_{\alpha,-\theta,\gamma,\kappa}(-\xi)$. Therefore, we formally leave $\theta$ undetermined, even if we assume only extremal cases, for which $\beta=-1$.

According to \cite{Diethelm, Podlubny}, the inverse Laplace transform of Eq.~\eqref{eq: diffrepr} is equal to the Mittag-Leffler function
 \begin{equation}\label{eq: gpt}
\hat{g}(p,\tau) = \tau^{\kappa-1} E_{\gamma,\kappa}\left(\mathcal{H}_{\alpha,\theta}(p) \tau^\gamma \right)\, .
\end{equation}
The Mittag-Leffler function can be represented through the Mellin integral transform which is defined as
\begin{equation}
\mathcal{M}[g(x);s] = \int_0^\infty g(x) x^{s-1} \ud x\,
\end{equation}
and the inverse transform is defined as (for some $c$ given by the Mellin transform theorem \cite{Flajolet})
\begin{equation}
g(x) = \frac{1}{2 \pi i} \int_{c-i \infty}^{c+ i \infty} \mathcal{M}[g](s) x^{-s} \ud s\, .
\end{equation}
This representation can provide the Green function in a more tractable form which can be better exploited in numerical computations. The Mittag-Leffler function can be expressed as a complex integral
\begin{equation}
E_{a,b}(z) = \frac{1}{2 \pi i} \int_{c - i \infty}^{c+i \infty} \frac{\Gamma(s') \Gamma(1-s')}{\Gamma(b- a s')} (-z)^{-s'} \ud s'
\end{equation}
where $0 < \Re(c) < 1$. Plugging into the equation~\eqref{eq: gpt}, we find that
\begin{equation}
\hat{g}(p,\tau) = \frac{\tau^{\kappa-1}}{2 \pi i} \int_{c - i \infty}^{c+i \infty} \frac{\Gamma(s')\Gamma(1-s')}{\Gamma(\kappa-\gamma s')} \left[ -\mu |p|^\alpha exp\left(-\frac{i \pi \theta \ \mathrm{sign}(p)}{2}\right) \tau^\gamma\right]^{-s'} \ud s'\, .
\end{equation}
Transforming the variable $p$ back to variable $\xi$, we obtain
\begin{equation}
g(\xi,\tau) = \frac{\tau^{\kappa-1}}{2 \pi i |\xi|} \int_{c - i \infty}^{c+i \infty} \frac{\Gamma(s')\Gamma(1-s')}{\Gamma(\kappa-\gamma s') \Gamma(s' \alpha)} \left[-\mu \frac{\tau^\gamma}{\xi^\alpha}\right]^{-s'} \ud s'\, .
\end{equation}
After change of variables $\alpha s' = s$ and taking into account the normalization \eqref{eq: riesz}, which can be in both cases written as $\frac{\tau^{\kappa-1}}{\Gamma(\kappa)}$, we finally arrive at the normalized Green function
\begin{equation}\label{eq: dfg}
g^{DF}(\xi,\tau) = \frac{\Gamma(\kappa)}{2 \alpha \pi i \xi}  \int_{c - i \infty}^{c + i \infty} \frac{\Gamma\left(\frac{s}{\alpha}\right) \Gamma\left(1-\frac{s}{\alpha}\right)\Gamma(1-s) }{\Gamma\left(\kappa-\frac{\gamma}{\alpha} s\right)\Gamma\left(\frac{(\alpha-\theta)s}{2\alpha} \right)\Gamma\left(1- \frac{(\alpha-\theta)s}{2\alpha}\right)} \left[\frac{\xi}{(-\mu \tau^{\gamma})^{1/\alpha}} \right]^{s} \ud s.
\end{equation}
From Eq.~\eqref{eq: dfg} it is apparent that $g(\xi,\tau) = \frac{1}{\tau^{\gamma/\alpha}} g\left(\frac{\xi}{\tau^{\gamma/\alpha}},1\right)$, so the Green function has the expected scaling with the temporal scaling exponent equal to $\frac{\gamma}{\alpha}$. The ratio $\Omega = \frac{\gamma}{\alpha}$ is called the \emph{diffusion scaling exponent}. Differences between Riesz-Feller and Caputo Green functions for various values of $\alpha$ and $\gamma$ are displayed in Fig.~\ref{fig: green}.

\section{Double-fractional Option Pricing Model}\label{sec: dfop}
%
\begin{figure}[t]
\includegraphics[width=5cm]{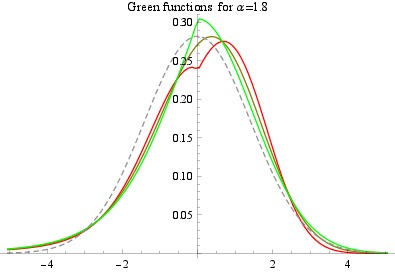} \quad
\includegraphics[width=5cm]{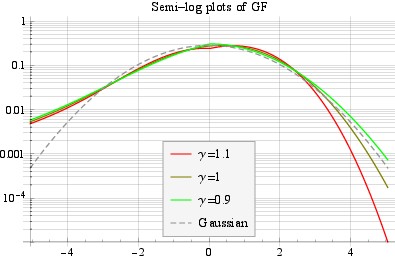} \quad
\includegraphics[width=5cm]{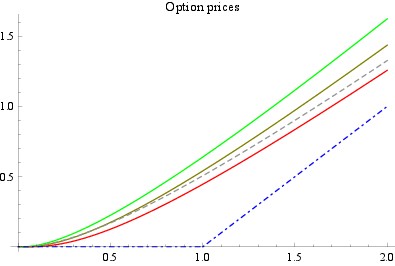}\\
\includegraphics[width=5cm]{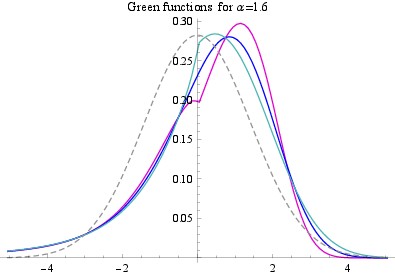} \quad
\includegraphics[width=5cm]{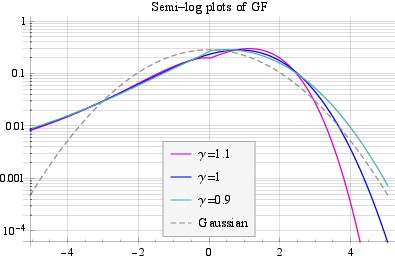} \quad
\includegraphics[width=5cm]{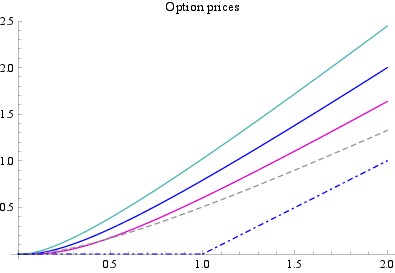}
\caption{Green functions (left), semi-log plots of $g^{DF}(x)$ (center) and corresponding option prices (right) for $\alpha=1.8$ (top line) and $\alpha=1.6$ (bottom line) and comparison with the Black-Scholes model (grey dashed lines). For some particular choices of parameters, there exist regions, where the option prices are cheaper than the Black-Scholes model and vice versa. The parameters are set to $\sigma=1$ and $\tau=1$.}
\label{fig: option}
\end{figure}
The solution of option-pricing equation driven by double-fractional diffusion under interest rate $r$ and dividend yield $q$ can be obtained by integrating over all scenarios $[S_T - K]^+$, so it reads
\begin{equation*}
C_{(\alpha,\gamma,\kappa)}(S_t,K,\tau) = e^{-r \tau} \int_\mathds{R} \ud y \ \left[S_t e^{\tau(r-q+\mu)+y} - K\right]^+ g^{DF} (y,\tau) = \end{equation*}
\begin{equation}
= e^{-r \tau} \int_\mathds{R} \ud y \ \left[S_t e^{\tau(r-q+\mu)+y} - K\right]^+ \frac{\Gamma(\kappa)}{2 \alpha \pi i y}  \int_{c - i \infty}^{c + i \infty} \frac{\Gamma\left(\frac{s}{\alpha}\right) \Gamma\left(1-\frac{s}{\alpha}\right)\Gamma(1-s) }{\Gamma\left(\kappa-\frac{\gamma}{\alpha} s\right)\Gamma\left(\frac{(\alpha-\theta)s}{2\alpha} \right)\Gamma\left(1- \frac{(\alpha-\theta)s}{2\alpha}\right)} \left[\frac{y}{(-\mu \tau^{\gamma})^{1/\alpha}} \right]^{s} \ud s.
\end{equation}
In Fig.~\ref{fig: option}  Green functions for different pairs $(\alpha,\gamma)$ and semi-log plots are displayed for better tail-behavior illustration. Option prices derived from Green functions are also presented. The corresponding put price can be obtained through put-call parity relation, which reads:
\begin{equation}
P_{(\alpha,\gamma,\kappa)}(S_t,K,\tau) = C_{(\alpha,\gamma,\kappa)}(S_t,K,\tau) - S_t  e^{- q \tau} + K e^{-r \tau}.
\end{equation}

\subsection{Model calibration for S\&P 500 options traded in November 2008}
\begin{figure}[t]
\includegraphics[width=6cm]{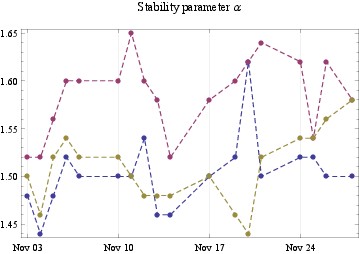} \quad
\includegraphics[width=6cm]{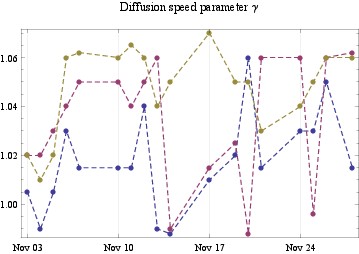} \\
\includegraphics[width=6cm]{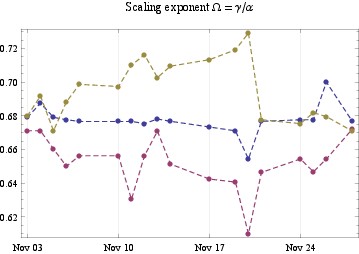} \quad
\includegraphics[width=6cm]{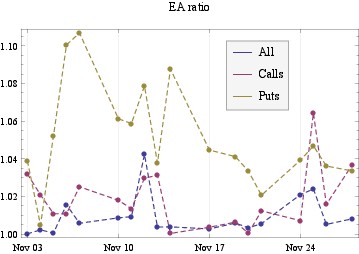}
\caption{Estimated values of stability parameter $\alpha$, diffusion speed parameter $\gamma$, scaling exponent $\Omega$ and the ratio $\rho$ of aggregated errors between L\'{e}vy model and Double-fractional model for each particular day, done for all options, resp. for calls and puts separately. We can observe that in case of calls and puts the benefit of the Double-fractional model is more significant. In the case when we observe a drop in $\alpha$ parameter, we can simultaneously observe a drop in values of $\gamma$, which points to the risk redistribution from long-term to short-term options. This phenomenon could be an indicator of market regime change. The parameter $\Omega$ measures the ratio between $\gamma$ and $\alpha$ and corresponds to the temporal scaling exponent, so $g(\xi,\tau) \sim \tau^\Omega$. We shall note that for BS model $\Omega=\frac{1}{2}$, which corresponds to Brownian motion. The graph shows that $\Omega$ exhibits more stable behavior, especially when estimated on both put and call options simultaneously.}
\label{fig: fit}
\end{figure}

\begin{table}[t]
\begin{tabular}{|c|ccc|}
  \hline
  \multicolumn{4}{|c|} {All options}\\
  \hline
 parameter& Black-Scholes & L\'{e}vy stable & Double-fractional  \\
\hline
  $\alpha$ & - & 1.493(0.028)& 1.503(0.037)\\
  $\gamma$  & - & - & 1.017(0.019) \\
  $\sigma$ & 0.1696(0.027)& 0.140(0.021)& 0.143(0.030) \\
   AE & 8240(638)& 6994(545)& 6931(553)\\
  \hline
  \hline
  \multicolumn{4}{|c|} {Call options}\\
  \hline
 parameter& Black-Scholes & L\'{e}vy stable & Double-fractional  \\
\hline
  $\alpha$ & - & 1.563(0.041)& 1.585(0.038)\\
  $\gamma$  & - & - & 1.034(0.024) \\
  $\sigma$ & 0.140(0.021) & 0.118(0.026) & 0.137(0.020)  \\
   AE & 3882(807) & 3610(812) & 3550(828) \\
  \hline
  \hline
    \multicolumn{4}{|c|} {Put options}\\
  \hline
 parameter& Black-Scholes & L\'{e}vy stable & Double-fractional  \\
\hline
  $\alpha$ & - & 1.493(0.031)& 1.508(0.036)\\
  $\gamma$ & - & - & 1.047(0.017)\\
  $\sigma$ & 0.193(0.039) & 0.163(0.034) & 0.163(0.037)   \\
  AE & 3741(711) & 3114(591) & 2968(594)\\
  \hline
\end{tabular}
\caption{Estimated mean values and standard deviations of model parameters $(\alpha,\gamma,\sigma)$ and mean aggregated error (MAE) for three considered models, i.e., Black-Scholes model, L\'{e}vy stable model with ordinary time derivative and Double-fractional model. The statistics are calculated for all options and separately for call options and put options. Apparently, the mean value of $\gamma$ is very close to one for all options. But when call options and put options are analyzed separately, the $\gamma$-value is larger than one. The mean square error of Double-fractional model exhibits significant improvement with respect to the Black-Scholes model, but compared with L\'{e}vy stable model, it shows only a little improvement. This is caused by the fact, that the assumption of constant parameters does not fully describe the complex behavior of option markets. The situation improves if we calibrate the model for call and put options separately. In Fig.~\ref{fig: fit} it is shown that on some particular days, the improvement of aggregated error is more significant.}
\label{tab: par}
\end{table}
We calibrate our model on the data of S\&P 500 options that were traded during November 2008. The choice of this period is mainly because of the financial crisis, which brought about phenomena that are  potentially interesting. We follow the methodology of Carr and Wu~\cite{Carr} and try to find such a triplet $(\alpha,\gamma,\sigma)$ that minimizes the aggregated option price error
\begin{equation}
AE_{\mathrm{model}} = \sum_{\tau \in \mathcal{T},K \in \mathcal{K}} | \mathcal{O}_{\mathrm{model}} - \mathcal{O}_{\mathrm{market}}|\,
\end{equation}
of all out-of-the money options, so
\begin{equation}
(\alpha_O,\gamma_O,\sigma_O) =  \mathrm{arg} \min_{(\alpha,\gamma,\sigma)} AE_{DF}(\alpha,\gamma,\sigma)\, .
\end{equation}
We make the optimization for each trading day. We have chosen the out-of-the-money options, because in-the-money option prices are more determined by the boundary conditions in option pricing formula, rather than by particular underlying diffusion model~\cite{Carr}. The statistics of calibrated parameters is listed in Tab.~\ref{tab: par} for the Black-scholes model, L\'{e}vy stable model and Double-fractional model. Because of the fact that $\gamma$ was close to $1$, the choice of derivative type did not have a large impact on the solutions and the results are practically the same, therefore the results are presented only for Caputo derivative, i.e., $\kappa=1$. The parameter $\alpha$ is close to $1.5$ in all cases, only for call options it fluctuates around $1.6$. The analysis shows that the double-fractional model brings more improvement when it is fitted separately for call and put options. This could be connected to the discussion about the general validity of put-call parity and market efficiency~\cite{Brunetti}. Fig.~\ref{fig: fit} shows estimated parameters for every day and also the ratio between aggregated errors of L\'{e}vy stable model and Double-fractional model $\rho = \frac{AE_{LS}}{AE_{DF}}$. Particularly for put options we can observe a noticeable improvement. Additionally,  another interesting phenomenon connected with the decrease of $\alpha$ parameter can be observed. In such a situation  the $\gamma$ parameter also decreases, even to values smaller than one, which could be interpreted as the risk redistribution from long-term options to short term options. The parameter $\Omega$, which is the ratio between parameters $\gamma$ and $\alpha$, expresses the temporal scaling parameter of the system. This parameter produces more stable behavior, pointing to the simultaneous changes in both parameters $\gamma$ and $\alpha$. The differences between option prices in the case of Double-fractional and Black-Scholes model are presented in Fig.~\ref{fig: prices}. One can observe that the price differences are relatively small, even in some cases are Double-fractional prices below prices estimated by the \mbox{Black-Scholes} model. This is caused by the fact that the Double-fractional Green function redistributes the probability of particular scenarios. In cases, when the option execution is less probable than in case of Black-Scholes model, we obtain cheaper option price. Nevertheless, with increasing maturity time  the Double-fractional prices become more expensive, but the difference is still not dramatic. We discuss the topic of risk redistribution and hedging in the next section.

\subsection{Risk Redistribution and Hedging Policy under Double-fractional Diffusion Model}
We notice a few properties of option prices driven by double-fractional diffusion. Indeed, we recover classical BS model for $\alpha=2$ and $\gamma=1$. When $\alpha<2$, the underlying return distribution is skewed and we have polynomial decay for the negative tail of the distribution. This results in an adjustment of the option price. A greater probability of fall of the underlying asset price results in an increase of option price for $K < F$~\footnote{$F$ is the forward price, i.e., $F = S_t \exp[(r-q)\tau]$.}, and options, for which $K > F$, become cheaper (both puts and calls). Similarly, the parameter $\gamma$ plays analogous role in temporal risk redistribution. For $\gamma < 1$, options with short expiration period become more expensive, while options with long expiration period become slightly cheaper. This behavior can be observed in situations when we face some kind of unexpected or sudden change of regime, such as a black day on the market, the bankruptcy of a company trading on the market, a natural disaster, etc. Indeed, for options with long expiration long-term equilibrium volatility is more important. Nevertheless, for options with short expiration such jumps and short-term uncertainty are the most important factors for price estimation. On the other hand, for $\gamma > 1$, the diffusion is faster than in case of space-fractional diffusion, and the options with long maturity time.
We should note that the change of parameters $\alpha$ or $\gamma$ does not change all option prices in the same direction; there are always options which become cheaper and more expensive. This essential role is played by the parameter $\sigma$, which is the volatility of the system.

\begin{figure}[t]
\includegraphics[width=6cm]{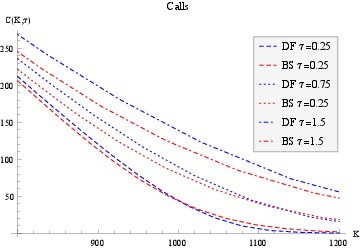} \quad
\includegraphics[width=6cm]{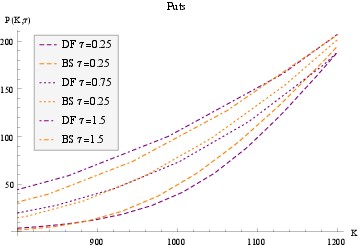} \\
\includegraphics[width=6cm]{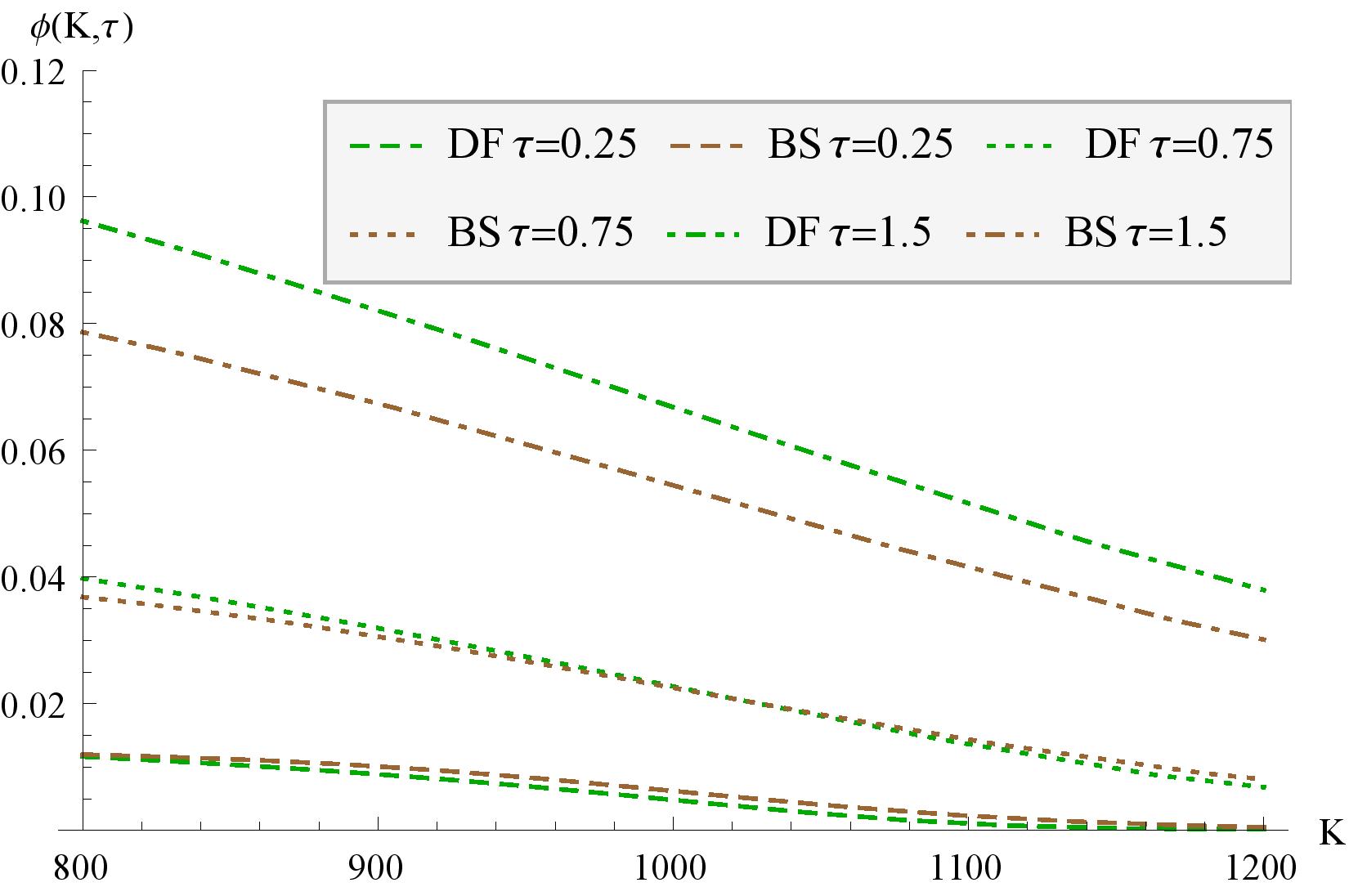} \quad
\includegraphics[width=6cm]{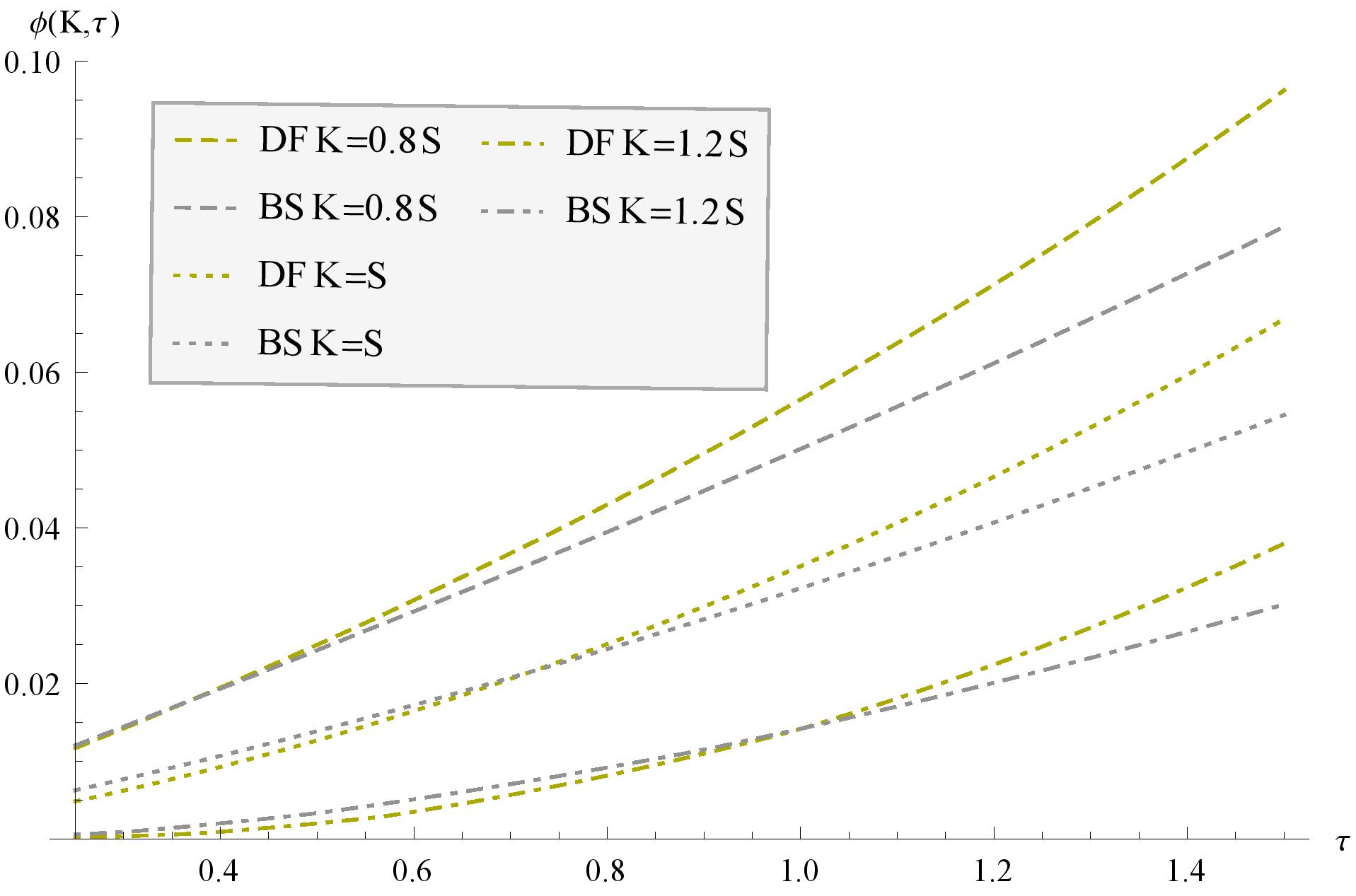}
\caption{Top: Estimated call and put prices of S\&P 500 calibrated by Double-fractional (DF) model and Black-Scholes (BS) model as functions of strike price $K$ for several maturity times $\tau$. The market spot price is $S=1000$. The model parameters are listed in Tab.~\ref{tab: par}. One can observe that the price differences are relatively small, increasing with increasing $\tau$, which is caused by different scaling exponents $\Omega$. In some cases  the price of DF option is even cheaper than in case of BS.\\
Bottom: Optimal policies $\phi^*(K,t)$ obtained from estimated parameters as functions of $K$, resp. $\tau$ for Double-fractional model and Black-Scholes $\Delta$-hedging. }
\label{fig: prices}
\end{figure}

The options are usually used for hedging the risk coming from the random nature of price evolution in financial markets. In the simplest case, with one underlying asset and one option, we use the \emph{$\phi$-hedging} strategy~\cite{Bouchard}. We create a portfolio of an option
\begin{equation}
\Pi(S,t) = C(S,t) - \phi(S,t) S(t)
\end{equation}
where $\phi(t)$ is the amount of stocks used to hedge the short position. There exist several risk measures. Nevertheless, we work with the most popular risk measure defined as variance of portfolio between times $t_0=0$ and $T$
\begin{equation}
\mathcal{R} = \langle (\Delta \Pi(S,t))^2 \rangle = \langle ([S-K]^+ - C(S_0,K,T) - \phi \Delta S)^2 \rangle
\end{equation}
The advantage of using this risk measure is that the optimal policy is easily expressible. The minimal risk is therefore determined by
\begin{equation}
\frac{\partial R}{\partial \phi} = 0\, .
\end{equation}
In Ref.~\cite{Busca} it was shown that the optimal hedging policy is given by
\begin{equation}
\phi^*(S,t) = \frac{1}{\sigma^2} \langle (S_0 - S) [S-K]^+ \rangle = \frac{1}{\sigma^2} \int_0^\infty \ud S (S_0 - S) [S-K]^+ g(S,T|S_0,t)
\end{equation}
where $\sigma^2$ is volatility of stock $S$. For Black-Scholes model we obtain the well-known $\Delta$-hedging rule
\begin{equation}
\phi^*_{BS}(S,t) = \Delta(S,t) = \frac{\partial C(S,t)}{\partial S}\, .
\end{equation}
The difference in optimal policies of Black-Scholes model ($\Delta$-hedging) and Double fractional model is shown in Fig.~\ref{fig: prices}. Due to the risk-redistribution, the resulting policies are also changed in order to minimize the risk in the environment with double-fractional diffusion. As with the prices, the risk is, in the case of changing parameter $\alpha$, redistributed in the spatial axis (prices) and in the case of parameter $\gamma$ in the temporal axis (time to maturity).

\section{Conclusions and Perspectives}
A novel method of option pricing was proposed on the basis of fully asymmetric Double-fractional diffusion and a special example was calibrated for S\&P 500 options traded in November 2008. The presence of power-law behavior in prices has been discussed before in case of cotton prices~\cite{Mandelbrot2} and options~\cite{Carr, Tankov}. There have been various alternative attempts to reduce the  risks in portfolios. Examples are provided by models based on regime switching~\cite{Calvet}, stochastic volatility~\cite{Heston}, jump processes~\cite{Tankov}, etc. Alternatively, it is possible to redistribute the risk by a temporally-fractional derivative, which, similar to spatial asymmetry, redistributes the risk to either the short-period options or long-term options. Such model enables one to treat the short-term differently; instant risk coming from contemporary fluctuations, jump corrections, etc., is treated differently from the long-term risk, which is determined by the slow volatility of the system. This also influences the optimal hedging policy.

While the long-term average behavior of markets tends to be similar in systems driven by ordinary, first time-derivative diffusion (i.e., diffusion speed parameter equal to one), an adjustment of diffusion speed parameter $\gamma$ to values $\gamma \neq 1$ can often describe the system better. Such fact coheres with the complex nature of financial markets and particularly option trading. Further investigations of the topic and its interrelation to other models will be the subject of future research.  Connection with regime switching models or interpretation of $\gamma$ parameter as a ``regime-switching'' time-dependent parameter and the conjunction with other models are questions of high importance and can possibly reveal additional potential of models based on double-fractional diffusion.

\section*{Acknowledgements}
Authors want to thank to Petr Jizba and V\'{a}clav Zatloukal for valuable discussions and Dunstan Clarke for corrections. J. K. acknowledges support from GA\v{C}R, Grant no. P402/12/J077 and Grant no. GA14-07983S.
\bibliographystyle{apsrev4-1}
\bibliography{references}
\appendix
\section{Comparison of Smearing Kernels for Riesz-Feller and Caputo Derivatives}\label{sec: appendix}
\begin{figure}
\begin{center}
\includegraphics[width= 10cm]{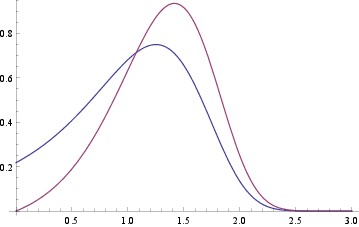}
\caption{Dependence of $g^{RF}(\tau,l)$ (purple line) and $g^{C}(\tau,l)$  (blue line) to the variable $l$ for $\tau=1$.}
\label{fig: ap}
\end{center}
\begin{picture}(0,0)
\put(0,33){$l$}
\put(-140,130){$g(\tau,l)$}
\end{picture}
\end{figure}
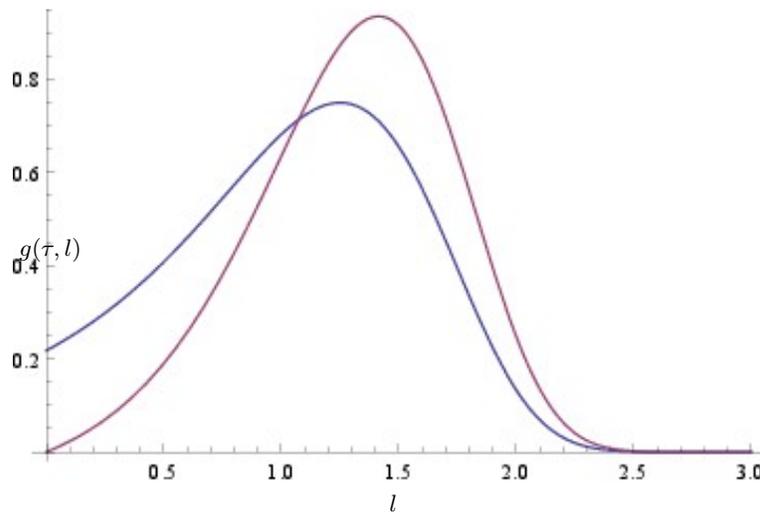
In this appendix, we compare Green function of Double-fractional diffusion in case when time derivative is equal to Riesz-Feller derivative and Caputo derivative and $\gamma<1$. The Green function is given by
\begin{equation}
g_{\alpha,\gamma}(\xi,\tau) = \int_0^\infty \ud l \, \psi_M(\tau,l) \frac{1}{l^{1/\gamma}} L_{\gamma,1}\left(\frac{\tau}{l^{1/\gamma}}\right) g_\alpha(\xi,l)\, ,
\end{equation}
where $\psi_M$ is different according to used derivative, so
\begin{equation}
\psi_M(l,\tau) = \left\{
                   \begin{array}{ll}
                     \frac{\Gamma(\gamma)}{\tau^{\gamma-1}} & \mathrm{for \ Riesz \ derivative}, \\
                     \frac{\tau}{l \gamma} & \mathrm{for \ Caputo \ derivative}.
                   \end{array}
                 \right.
\end{equation}
Indeed, it is interesting to see, what happens with the smearing kernel
\begin{equation}
g_{1,\gamma}(l,\tau) = \psi_M(\tau,l) \frac{1}{l^{1/\gamma}} L_{\gamma,1}\left(\frac{\tau}{l^{1/\gamma}}\right)
\end{equation}
for small and large values. Firstly, when $l \rightarrow 0$, then the argument of $\gamma$-stable distribution distribution goes to infinity. According to Ref.~\cite{Skorohod}, the asymptotic expansion gives us
\begin{equation}
\frac{1}{l^{1/\gamma}} L_{\gamma,1}\left(\frac{\tau}{l^{1/\gamma}} \right) \sim \frac{ \Gamma(\gamma+1) \sin(\pi \gamma)}{\cos \left( \frac{\pi \gamma}{2} \right)} \frac{l}{\tau^{\gamma+1}} \qquad \mathrm{for} \ l \rightarrow 0\, ,
\end{equation}
which gives us
\begin{equation}
g_1^{RF}(\tau,l) \sim \frac{l}{\tau^{2\gamma}} \frac{ \Gamma(\gamma) \Gamma(\gamma+1) \sin(\pi \gamma)}{\cos \left( \frac{\pi \gamma}{2} \right)} \qquad \mathrm{for} \ l \rightarrow 0.
\end{equation}
For Caputo case we get non-zero value of smearing kernel for $l=0$. Particularly, it is equal to
\begin{equation}
g_1^{C}(\tau,0) = \left(\frac{1}{\tau^\gamma}\right) \frac{\Gamma(\gamma) \sin(\pi \gamma)}{\cos \left( \frac{\pi \gamma}{2} \right)}\, .
\end{equation}
On the other hand, when $l \rightarrow \infty$, then it is necessary to use the Taylor expansion of $L_{\gamma,1}(x)$, and again from \cite{Skorohod} we obtain
\begin{equation}
L_{\gamma,1}(x) \sim A_\gamma x^{-1-\frac{\lambda_\gamma}{2}} \exp\left(-B_\gamma x^{-\lambda_\gamma}\right)\quad \mathrm{for} \ x \rightarrow 0^+\, ,
\end{equation}
where $\lambda_\gamma = \frac{\gamma}{1-\gamma}$ and $A_\gamma$ resp. $B_\gamma$ are $\gamma$-dependent constants. The asymptotic behavior can be therefore described as
\begin{equation}
g_1^{RF}(\tau,l) \sim C^{RF}(\tau) A_\gamma l^\frac{1}{2(1-\gamma)} \exp\left(- B_\gamma D(\tau) l^\frac{1}{1-\gamma} \right) \qquad \mathrm{for} \ l \rightarrow +\infty
\end{equation}
and for Caputo case as
\begin{equation}
g_1^{C}(\tau,l) \sim C^C(\tau) A_\gamma l^{\frac{1}{2(1-\gamma)}-1} \exp\left(- B_\gamma D(\tau) l^\frac{1}{1-\gamma} \right) \qquad \mathrm{for} \ l \rightarrow 0.
\end{equation}
 The $\tau$-dependent constants $C^{RF}(\tau)$, resp. $C^{C}(\tau)$ can be determined from previous expressions. Therefore, in both cases we become exponential decay in $l$. The graphs of both smearing kernels are depicted in Fig.~\ref{fig: ap}.

\end{document}